# ISOSPIN DIFFUSION IN HEAVY ION REACTIONS

M.B. Tsang, T.X. Liu, L. Shi, P. Danielewicz, C.K. Gelbke, X.D. Liu, W.G. Lynch, W.P. Tan[a], G. Verde, A. Wagner[b], H.S. Xu[c]

*National Superconducting Cyclotron Laboratory and Physics and Astronomy Department, Michigan State University, East Lansing, MI 48824, USA,*

W.A. Friedman,

*Department of Physics, University of Wisconsin, Madison, WI 53706, USA,*

L. Beaulieu[d], B. Davin, R.T. de Souza, Y. Larochelle[e], T. Lefort, R. Yanez, V.E. Viola Jr.

*Department of Chemistry and IUCF, Indiana University, Bloomington, IN 47405, USA,*

R.J. Charity, and L.G. Sobotka,

*Department of Chemistry, Washington University, St. Louis, MO 63130, USA*

Using symmetric $^{112}$Sn+$^{112}$Sn, $^{124}$Sn+$^{124}$Sn collisions as references, we probe isospin diffusion in peripheral asymmetric $^{112}$Sn+$^{124}$Sn, $^{124}$Sn+$^{112}$Sn systems at incident energy of E/A=50 MeV. Isoscaling analyses imply that the quasi-projectile and quasi-target in these collisions do not achieve isospin equilibrium, permitting an assessment of the isospin transport rates. We find that comparisons between isospin sensitive experimental and theoretical observables, using suitably chosen scaled ratios, permit investigation of the density dependence of the asymmetry term of the nuclear equation of state.

---

[a] Present address: Dept of Physics, Notre Dame University, IN.
[b] Present address: Forschungszentrum Rossendorf e.V., Dresden, Germany.
[c] Present address: Institute of Modern Physics, CAS, Lanzhou, China.
[d] Present address: Université Laval, Quebec City, Canada, G1K-7P4
[e] Present address: Oak Ridge National Laboratory, Oak Ridge TN.

The nuclear mean field potential binds nuclei, stabilizes neutron stars against gravitational collapse [1] and generates forces that shape the dynamics of nuclear collisions [2-4] and supernova explosions [5]. Many of these environments involve densities or isospin-asymmetries far from those characteristic for stable nuclei [1,5]. Investigations of nucleus-nucleus collisions enable experimental constraints to be placed on the mean field potential and on nuclear equation of state (EOS) at the extremes [2-4,6,7]. Significant constraints have already been placed on the symmetric matter EOS at high densities [2]. In contrast, relatively weak constraints exist on the isospin-asymmetry term of the EOS, which describes the EOS sensitivity to the difference between neutron and proton densities and allows for extrapolations to the neutron-rich matter within neutron stars [1,8].

Comparisons of collisions of neutron-rich to that of neutron-deficient systems provide one means of probing the asymmetry term experimentally [9-11]. A second involves "isospin-asymmetric" collisions of projectile and target nuclei. In such cases, the asymmetry term of the mean field provides driving forces that propel the system towards isospin equilibrium where anisotropies in the difference between neutron and proton densities are minimized. Investigations of these driving forces require collisions with commensurate collision and isospin-equilibration timescales and observables that reveal isospin equilibration. At low incident energies, deep inelastic collisions have been used to study charge equilibrium leading to conflicting conclusions [12,13]. However at incident energy above E/A=30 MeV, the time scale for fragment emission decreases and anisotropies in the emission patterns may develop [14], which may allow one to measure from the fragments the time scales for charge and mass transport and diffusion during the collision. Here, we show that isotopic distributions near projectile rapidity in peripheral Sn+Sn collisions at E/A=50 MeV allow for investigations of isospin diffusion. Transport calculations suggest that sub-saturation density is achieved in the neck region where isospin diffusion occurs [15]. By comparing the results of measurements and calculations relevant to isospin diffusion, we explore the sensitivity of the diffusion to the density dependence of the asymmetry term of the EOS at sub-saturation densities.

The experiment involved bombarding $^{112}$Sn and $^{124}$Sn targets of 5 mg/cm$^2$ areal density with 50 MeV per nucleon $^{112}$Sn and $^{124}$Sn beams from the K1200 cyclotron at the National Superconducting Cyclotron Laboratory at Michigan State University [11]. Nine telescopes of the Large Area Silicon Strip Array (LASSA) [16] detected isotopically resolved particles with 1≤Z≤8. Each telescope consists of one 65 μm single-sided silicon strip detector, one 500 μm double-sided silicon strip detector and four 6-cm thick CsI(Tl) scintillators, read out by pin diodes. The strips of the silicon detectors divide the 50mm x 50mm lateral dimensions of each telescope into 256 (3x3 mm$^2$) square pixels, providing an angular resolution of about ±0.43°.

We placed the center of the LASSA device at the polar angle of θ=32° with respect to the beam axis, covering polar angles of 7°≤ θ ≤ 58°. Additionally, 188 plastic scintillator - CsI(Tl) phoswich detectors of the Miniball/Miniwall array [17] complemented the LASSA coverage. The combined apparatus covered 80% of the total solid angle. Peripheral collisions were selected by gates on the charged-particle multiplicity yielding the reduced impact parameter range of $b/b_{max}$ = 0.8-1.0 in the sharp cut-off approximation [18]. To minimize contributions from the neck fragments, projectile rapidity fragments were selected by the rapidity gate of $y/y_{beam}$ ≥ 0.7, where y and $y_{beam}$ are rapidities of the analyzed particle and beam, respectively. The experimental results are the same within the statistical uncertainty if a more restrictive rapidity gate of $y/y_{beam}$ ≥ 0.8 is used instead.

Because there are no isospin differences between identical projectiles and targets, the symmetric $^{112}$Sn +$^{112}$Sn and $^{124}$Sn +$^{124}$Sn collisions are used to establish diffusion-free baseline values for the measured and predicted observables. The asymmetric $^{124}$Sn +$^{112}$Sn and $^{112}$Sn +$^{124}$Sn collisions, on the other hand, have the large isospin differences needed to explore the isospin diffusion. As experimental observables, we focus on features of the isotopic yields $Y_i$(N,Z) of particles measured for reaction "i" at $y/y_{beam}$ ≥ 0.7. Here, N and Z are the neutron and proton numbers for the detected particles. As in multi-fragmentation [9,10,19], in deep inelastic collisions [10,20] and in statistical evaporation [10], we find that ratios of isotopic yields $R_{21}$(N,Z)=$Y_2$(N,Z)/$Y_1$(N,Z) for a specific pair of reactions, with a different total isotopic composition, follow the isoscaling relationship [10]

$$R_{21}(N,Z)=Y_2(N,Z)/Y_1(N,Z)= C\exp(\alpha N+\beta Z), \quad (1)$$

where α and β are the isoscaling parameters for the chosen reactions pair.

The left panel of Fig. 1 displays the measured values for $R_{21}(N,Z)$ when using $^{124}$Sn +$^{124}$Sn collisions as reaction 2 and $^{112}$Sn+$^{112}$Sn collisions as reaction 1 in Eq. 1. The fits following the right hand side of Eq. 1 are represented by the solid and dashed lines. Fits to the other two systems relative to $^{112}$Sn+$^{112}$Sn (not shown) are of similar quality. In the right panel of Fig. 1, we plot the best fit values for the isoscaling parameter, $\alpha$, versus the overall isospin asymmetry of the colliding system: $\delta_O = (N_O-Z_O)/(N_O+Z_O)$, where $N_O$ and $Z_O$ are the corresponding total neutron and proton numbers. The solid and open points represent data for $^{124}$Sn and $^{112}$Sn projectiles, respectively. In general, the isoscaling parameter $\alpha$ increases with the overall isospin asymmetry $\delta_O$. If a linear correlation between $\alpha$ and $\delta_O$ is assumed [11], then the mixed systems at $\delta_O = 0.153$ should assume an $\alpha$ value midway between that of $^{124}$Sn+$^{124}$Sn ($\delta_O = 0.193$, $\alpha=0.57\pm0.02$) and $^{112}$Sn+$^{112}$Sn ($\delta_O = 0.107$, $\alpha=0$) if isospin is equilibrated across the system. However, the measured value for the $^{124}$Sn projectile (solid point) is much larger than the value for the $^{112}$Sn projectile (open point) at $\delta_O = 0.153$ indicating that isospin equilibrium is not achieved in the asymmetric reaction systems [21]. The upper value of $\alpha=0.42\pm0.02$, obtained for $^{124}$Sn+$^{112}$Sn, represents, in the linear relation, an effective asymmetry of about 0.17. This value corresponds to roughly half way from the projectile value of 0.193 to the "equilibrium" value of 0.153. The lower value of $0.16\pm0.02$, obtained for $^{112}$Sn+$^{124}$Sn, has the same interpretation; except here the projectile is $^{112}$Sn and the change in asymmetry is in the opposite direction. The consistency of these results extracted from two independent measurements adds credibility to our approach.

Following ref. [22], we define the isospin transport ratio $R_i$ as

$$R_i = \frac{2x - x_{124+124} - x_{112+112}}{x_{124+124} - x_{112+112}} \qquad (2)$$

where $x$ is an isospin sensitive observable, preferably linear in asymmetry. For the two symmetric systems $^{124}$Sn+$^{124}$Sn and $^{112}$Sn+$^{112}$Sn, $R_i$ is automatically normalized to $+1$ and $-1$, respectively, allowing for quantitative comparison of the measured and predicted $R_i$ values even if the model calculations use isospin observables that differ from the experimental ones. In the absence of isospin diffusion, preequilibrium emission from the projectile should be approximately equal for asymmetric $^{124}$Sn+$^{112}$Sn ($^{112}$Sn+$^{124}$Sn) collisions as for symmetric $^{124}$Sn+$^{124}$Sn ($^{112}$Sn+$^{112}$Sn) collisions. By focusing on differences in isospin observables between mixed and symmetric systems $R_i(\alpha)$ largely removes the sensitivity to preequilibrium emission and enhances the sensitivity to diffusion. To represent the experimental data, we chose the neutron isoscaling parameter $\alpha$ because the projectiles and targets for the asymmetric systems differ only in their neutron numbers. Using $\alpha$ for $x$ in Eq. 2, we obtain the isospin transport ratios of the two asymmetric systems shown as the shaded bands in Fig. 2. The observed values, $|R_i(\alpha)| \approx 0.5$, are consistent with previous discussion that the isospin asymmetry of the projectile remnant is half way between that of the projectile and the "equilibration value".

We now explore the relationship between isospin diffusion, and the asymmetry term of the EOS within the context of the Boltzmann-Uehling-Uhlenbeck (BUU) [23,24] formalism, which calculates the time evolution of the colliding system using a self-consistent mean field. The isospin independent part of the mean field in these calculations is momentum independent and described by an incompressibility coefficient of K=210 MeV [15]. The interaction component of the asymmetry term in three sets of calculations provides a contribution to the symmetry energy per nucleon of the form $E_{sym,int}/A = C_{sym}(\rho/\rho_o)^\gamma$ where $C_{sym}$ is set to 12.125 MeV and $\gamma=2$, $\gamma=1$ and $\gamma=1/3$. Here, smaller values for $\gamma$ dictate a weaker density dependence for $E_{sym,int}$. The fourth set of calculations, referred to as SKM, uses an interaction asymmetry term providing $E_{sym,int}/A = 38.5 (\rho/\rho_o) - 21.0(\rho/\rho_o)^2$ [25] and has the weakest density dependence at $\rho \leq \rho_o$.

Calculations were performed for the $^{124}$Sn + $^{124}$Sn, $^{124}$Sn + $^{112}$Sn, $^{112}$Sn + $^{124}$Sn and $^{112}$Sn + $^{112}$Sn systems at an impact parameter of b=6 fm [15]. We employed ensembles of 800 test particles per calculation and we followed each calculation for an elapsed time of 216 fm/c. At this late time, the projectile and residues can be cleanly separated. Nonetheless, we require that all nucleons in the assigned regions to have density less than $0.05\rho_0$ and velocities more than half of the beam velocity in the center of mass, to be consistent with the experimental gates. To reduce statistical fluctuations in the results, we averaged them over 20 calculations for each system.

Using the average asymmetry of the projectile-like residue $\langle\delta\rangle$ in the calculation, as the isospin observable, $x$, in Eq. 2, we plot predictions for $R_i(\delta)$ as a function of time as bands in Fig. 3 for the stiffest asymmetry term, ($\rho^2$, top panel) and the softest asymmetry term, (SKM, bottom panel). By construction, $R_i(\delta)$ describes the evolution of isospin asymmetry for the projectiles ($^{112}$Sn or $^{124}$Sn) in the mixed reactions relative to that for the symmetric $^{124}$Sn + $^{124}$Sn (with $R_i$ =1) and $^{112}$Sn + $^{112}$Sn ($R_i$ =-1) systems. The widths of the bands reflect the statistical uncertainties of the calculated values for $R_i(\delta)$. Initially, these predictions for $R_i(\delta)$ represent the isospin of the projectile ($R_i(\delta)$ =1 for $^{124}$Sn and $R_i(\delta)$ =-1 for $^{112}$Sn) nuclei. Subsequent isospin diffusion drives the $R_i(\delta)$ values towards zero. Even though pre-equilibrium emission from the projectile remnants influences $\langle\delta\rangle$, $R_i(\delta)$ is not strongly modified because such pre-equilibrium emission is largely target independent and therefore cancelled in $R_i(\delta)$ by construction.

The influence of the asymmetry term depends on its magnitude at sub-saturation density [24,26]. For the top panel, $E_{sym,int}/A = C_{sym}(\rho/\rho_o)^2$ decreases rapidly at low density and becomes very small, leading to little isospin diffusion. For the bottom panel, $E_{sym,int}/A$ =38.5 $(\rho/\rho_o)$ - 21.0$(\rho/\rho_o)^2$ remains larger at low density, leading to stronger isospin diffusion and driving the residues to approximately the same isospin asymmetry. In both cases, the asymptotic value for $R_i(\delta)$ is first reached at around 100 fm/c when the two residues separate and cease exchanging nucleons as illustrated by the time evolution images of the collisions for the $^{124}$Sn +$^{124}$Sn system in Figure 3. This timescale is comparable to the collision timescale $\tau_{coll}$, which can be roughly (~20%) estimated by

$$\tau_{coll} \approx (4R_N+d)/v_{beam} \approx 80 \text{ fm/c}, \qquad (3)$$

where $R_N$, $v_{beam}$ and d are the nuclear radius, incident velocity, and the separation distance between the two nuclear surfaces at breakup.

Assuming that the experimental isoscaling relationships, shown in Fig. 1, reflect particle emission from the projectile remnants after 100 fm/c and that such emission can be described statistically, the calculated values for $R_i(\delta)$ may be easily related to the measured ones. In doing so, we take advantage of the nearly linear relationship between $\alpha$ and δ that has been shown to be valid for evaporation and for statistical multi-fragmentation of the remnants that emit the observed particles [27]:

$$\alpha \propto (\delta_2 - \delta_1)\left(1 - \frac{\delta_2 + \delta_1}{2}\right) \qquad (4)$$

where $\delta_1$ and $\delta_2$ are the asymmetries for the two systems involved in the isoscaling ratio [27]. Inserting Eq. 4 into the expression for $R_i(\alpha)$, one can show

$$R_i(\alpha) = R_i(\delta) - \frac{(\delta_i - \delta_{112+112})(\delta_i - \delta_{124+124})}{(\delta_{124+124} - \delta_{112+112})(1 - [\delta_{124+124} + \delta_{112+112}]/2)} \qquad (5)$$

Since the second term is negligible (<4%), we can use asymptotic values of <δ> around 216 fm/c, as $x$ in Eq. 2 to obtain $R_i(\delta)$ values for these two systems and compare them to the experimental values of $R_i(\alpha)$. The calculated values of $R_i(\delta)$, shown as open points in Fig. 2 in the order of an increasing "softness" from left to right indicate an increased isospin equilibrium for successively softer asymmetry terms. The diamond shaped points indicate the $R_i(\alpha)$ values obtained as in the experiment by decaying the residues obtained in BUU to fragments using the statistical model of ref. [27]. As expected, there is close agreement between the diamonds and the corresponding open points according to Eq. 5.

Figure 2 demonstrates the sensitivity of such observables to the asymmetry term of the EOS. The experimental $R_i(\alpha)$ values are closest to the predicted $R_i(\delta)$ values derived from the theoretical predictions for <δ> using the stiffest asymmetry term with $E_{sym,int}/A$ =$C_{sym}(\rho/\rho_o)^2$. This conclusion depends, however, on the assumption that the measured particles are produced after 100 fm/c when <δ> attains its asymptotic values. If the data include emission from earlier stages when δ is larger, a favorable comparison with calculations using softer asymmetry terms may be possible. Relevant determinations of the emission timescales for the detected particles are being explored [26] and may make the present conclusions regarding the density dependent asymmetry term more definitive.

In summary, we have observed the effects of isospin diffusion by investigating heavy ion collisions at E/A=50 MeV with comparable diffusion and collision timescales. Simulations using the BUU transport model predict that the isospin diffusion reflects driving forces arising from the asymmetry term of the EOS. The present comparisons suggest a better agreement with the stiff asymmetry term; however, more stringent constraints on the emission timescales for the measured particles are needed. The current version of BUU has no momentum dependence of the mean field which might interplay with the symmetry term. However, at incident energies around 50 MeV/u, few nucleons get accelerated to high momentum so the momentum dependence of the mean field need not be that important. Nonetheless, such effect should be explored in the future.

This work is supported by the National Science Foundation under Grant Nos. PHY-01-10253, PHY-0245009, PHY-00-70161 and by the DOE under grant numbers DE-FG02-87ER-40316.REFERENCES

1. J.M. Lattimer and M. Prakash, Ap. J., **550**, (2001) 426 and refs. therein.
2. Pawel Danielewicz, Roy Lacey and William G. Lynch, Science 298, 1592 (2002).
3. H.H. Gutbrod etal.,. Rep. Prog. Phys. 52, 1267 (1989).
4. C. Sturm et al., PRL 86, 39 (2001).
5. H.A. Bethe, Rev. Mod. Phys. 62, 801 (1990)
6. Bao-An Li, Phys. Rev. Lett. 88, 192701 (2002) and refs. therein.
7. M. Di Toro et al., Eur. Phys. Jour. A 13, 155 (2002) and refs. therein
8. "Isospin Physics in Heavy-Ion Collisions at Intermediate Energies",Eds. Bao-An Li and W. Udo Schroeder, NOVA Science Publishers, Inc. (New York), (2001).
9. W. P. Tan et al., Phys. Rev. C 64, 051901(R) (2001).
10. M.B. Tsang et al., Phys. Rev. Lett. 86, 5023 (2001).
11. H. Xu et al., Phys. Rev. Lett. **85,** 716 (2000).
12. B. Gatty et al, Nuc. Phys. A253, 511 (1975).
13. R. Planeta et al Phys. Rev. C 38, 195 (1988), R. Planeta et al., Phys. Rev. C 41, 942 (1990).
14. S. J. Yennello et al., Phys. Lett. B321, 15 (1994).
15. L. Shi, Thesis, Michigan State University, (2003).
16. A. Wagner et. al., Nucl. Instr. & Meth. A456, 290 (2001); B. Davin et al., Nucl. Instr. & Meth. A**473**, 302 (2001)
17. R.T. de Souza et al. Nucl. Inst. Meth. Phys. Res. A295, 109 (1990).
18. L. Phair et al., Nucl. Phys. A **548**, 489 (1992).
19. D.V. Shetty, et al., Phys. Rev. C 68, 021602 (2003).
20. G.A. Souliotis, et al., Phys. Rev. C 68, 024605 (2003).
21. J. F. Dempsey et al., Phys. Rev, C **54**, 1710 (1996) and refs. Therein.
22. F. Rami et al., Phys. Rev. Lett. 84, 1120 (2000).
23. G. Bertsch and S. Das Gupta, Phys. Rep. 160, 189 (1988).
24. Bao-An Li, et al., Int. J. Mod. Phys. E **7** (1998) 147, and refs. therein.
25. V. Baran, et al., Nucl. Phys. A **632**, 287 (1998).
26. L. Shi and P. Danielewicz, nucl-th/0304030.
27. M. B. Tsang et al., Phys. Rev. C **64**, 054615 (2001)8

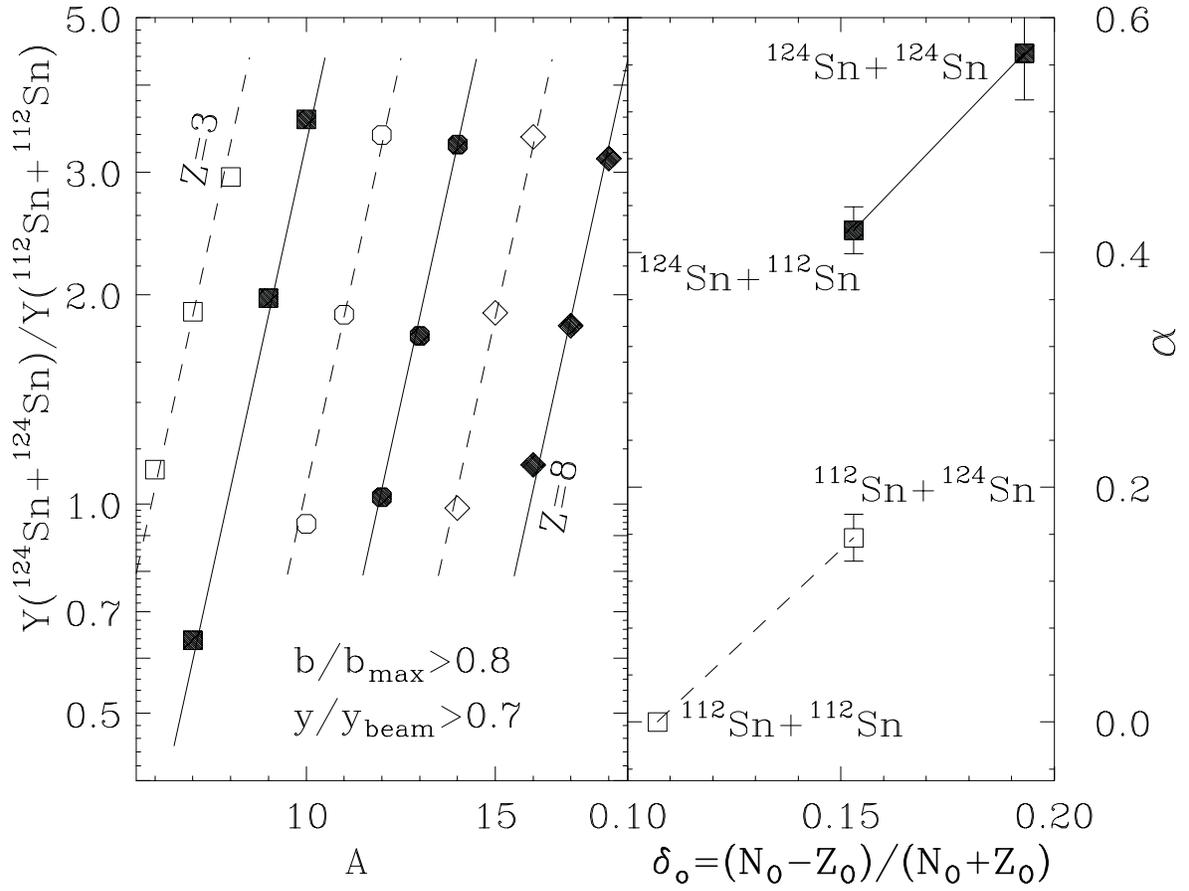

Fig. 1: Left panel: Measured values for $R_{21}(N,Z) = Y_{124+124}(N,Z)/Y_{112+112}(N,Z)$ (points) and fits with Eq. 1 (solid lines). The solid line and points represent even Z=4,6,8 isotopes while the dash lines and open points represent odd Z=3,5,7 isotopes. Right panel: Best fit values for $\alpha$ as function of $\delta_O$. The lines serve to guide the eye. The reactions are labelled next to the data points. Solid points denote $^{124}$Sn as the projectile and open points denote $^{112}$Sn as the projectile.

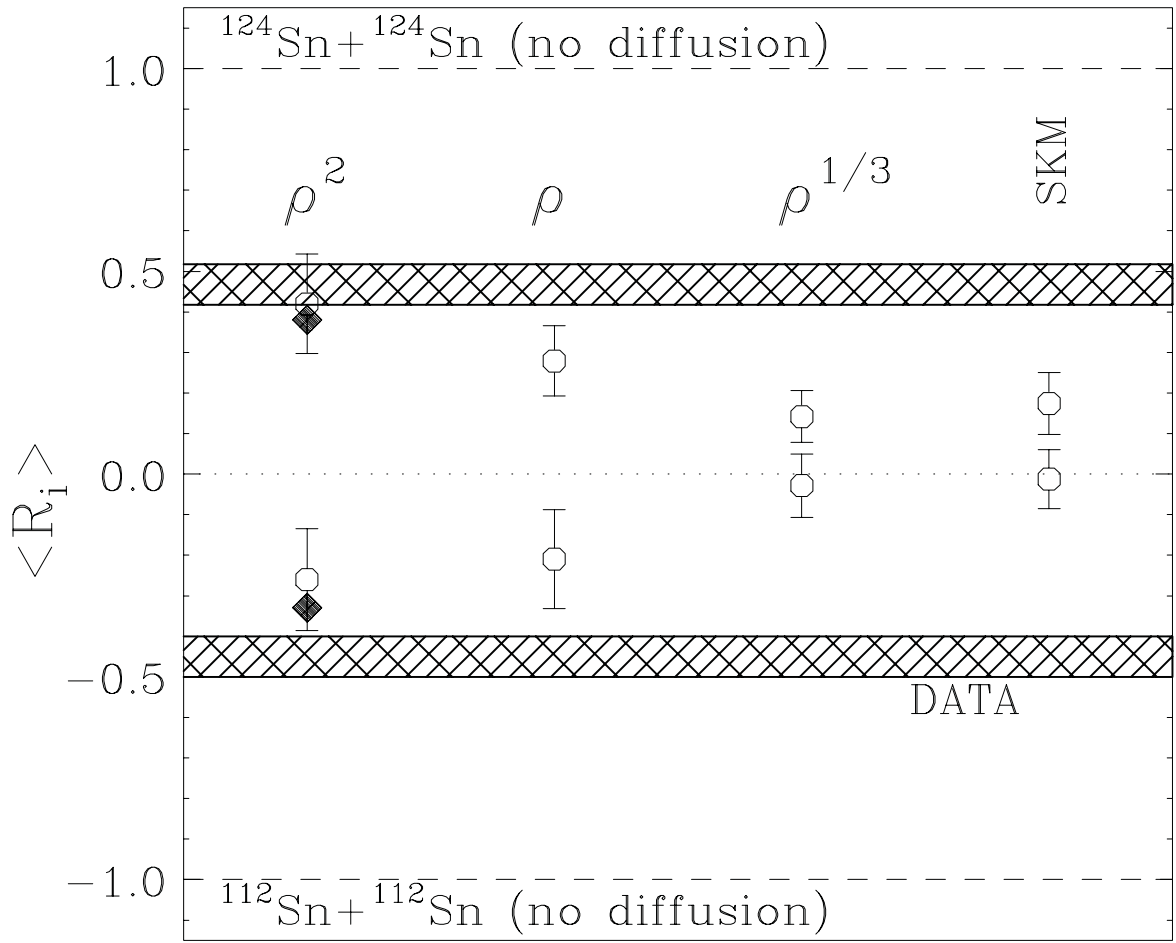

Fig.2: Measured (shaded bars) and calculated (open points) values for $R_i$. The labels on the calculated values represent the density dependence of $E_{sym,int}/A$ with increasing "softness" from left to right.

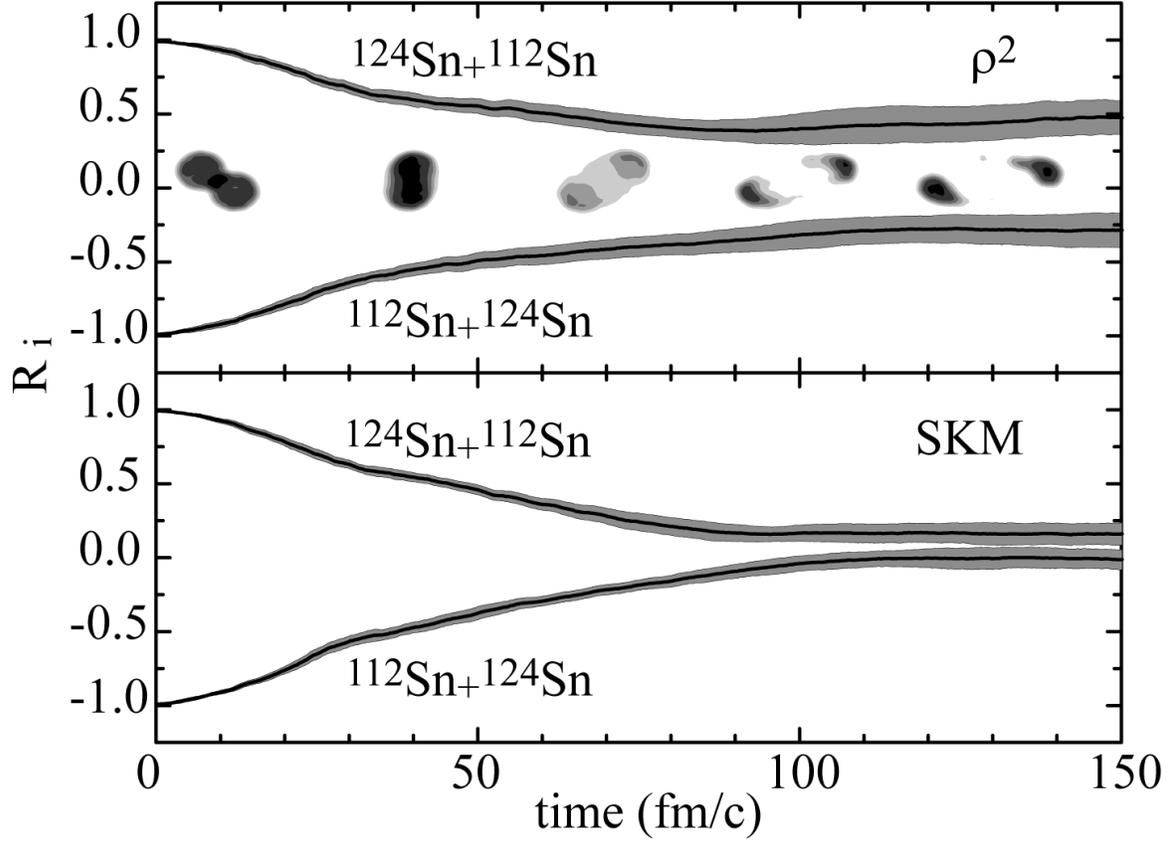

Fig. 3: $R_i(\delta)$ from BUU calculations are plotted as a function of time for the mixed systems. The symmetric systems are calibrated to +1 and −1 automatically by Eq. 2. The top and bottom panels show the calculated results with $E_{sym,int}/A = C_{sym}(\rho/\rho_o)^2$ and $E_{sym,int}/A = 38.5 (\rho/\rho_o) - 21.0(\rho/\rho_o)^2$, respectively. The bands above (below) zero represent the system with $^{124}$Sn ($^{112}$Sn) as the projectile. The time evolution images of the collisions for the $^{124}$Sn + $^{124}$Sn systems are superimposed on the upper panel suggesting that the projectile and target separate around 100 fm/c.